\documentclass[preprint,12pt]{elsarticle}

\usepackage{amssymb}
\usepackage{amsmath}
\usepackage{amsfonts}
\usepackage{bm}
\usepackage{graphicx}
\usepackage{hyperref}
\usepackage{enumitem}
\usepackage{url}

\journal{Computer Physics Communications}

\begin{document}

\begin{frontmatter}

\title{\texttt{Hubb\_DMFT} and \texttt{Wan2mb\_DMFT}: 
Continuous-Time Quantum Monte Carlo Solvers for Single-
and Multi-Orbital Hubbard Model}

\author{Andrey A. Katanin\corref{author}}
\cortext[author]{Corresponding author.}
\ead{Andrey.Katanin@gmail.com}
\affiliation{Center for Photonics and 2D Materials, Moscow Institute of Physics and Technology, Institutsky lane 9, Dolgoprudny, 141700, Moscow Region, Russia}
\affiliation{M. N. Mikheev Institute of Metal Physics of Ural Branch of Russian Academy of Sciences, S. Kovalevskaya Street 18, 620990 Yekaterinburg, Russia}


\begin{abstract}
We present two related 
customized software packages, \texttt{Hubb\_DMFT} and  \texttt{Wan2mb\_\linebreak[0]DMFT}, designed to solve the Dynamical Mean-Field Theory (DMFT) equations for strongly correlated electron systems. \texttt{Hubb\_DMFT} is adjusted for the single-band Hubbard model, providing a fast way to calculate the local self-energy, as well as the two-particle fermion and triangular fermion-boson charge and spin vertices, while \texttt{Wan2mb\_DMFT} extends this capability to realistic multi-orbital systems, directly interfacing Wannier tight-binding Hamiltonians with many-body solvers. Both codes are based on iQIST v.0.7 impurity solver \cite{iQIST1} and utilize a  modified and internally integrated Continuous-Time Quantum Monte Carlo (CT-QMC) core with the hybridization expansion (CT-HYB) framework and improved self-energy and vertex estimators for density-density interaction, ensuring numerically exact solutions for quantum impurity problems.\\
\\
\noindent {\bf PROGRAM SUMMARY:}\\
\begin{small}
  {\em Program Titles:} {Hubb\_DMFT} / {Wan2mb\_DMFT}\\
  {\em Developer's repository link:} During the pre-publication period provided upon request. \\
  {\em Licensing provisions:} GPLv3\\
  {\em Programming language:} Fortran (Core solvers and DMFT loops)\\
  {\em Nature of problem:} Solving the quantum impurity problem embedded in a self-consistent electronic bath for single- and multi-orbital lattice systems.\\
  {\em Solution method:} Continuous-Time Hybridization Expansion Quantum Monte Carlo (CT-HYB) algorithm combined with DMFT self-consistency loops.
	
\end{small}

\end{abstract}
\end{frontmatter}

\begin{keyword}
Dynamical Mean-Field Theory (DMFT) \sep Continuous-Time Quantum Monte Carlo (CT-QMC) \sep iQIST codebase \sep Hubbard Model \sep Wannier Functions \sep Hybrid Parallelization.
\end{keyword}

\section{Introduction}
Dynamical Mean-Field Theory (DMFT) \cite{DMFT} has revolutionized the understanding of strongly correlated electron systems by mapping a complex lattice problem onto a local quantum impurity model embedded in a self-consistent electronic bath. The predictive accuracy of DMFT heavily relies on the choice of the quantum impurity solver. Continuous-Time Quantum Monte Carlo (CT-QMC) methods \cite{CT-QMC}, specifically the hybridization expansion (CT-HYB) variant \cite{CT-QMC1}, provide a numerically exact solution without Trotter-Suzuki time-discretization errors. 

In this paper, we introduce a unified suite of two Fortran packages designed to address different regimes of strong correlations. Despite presence of many packages, solving quantum impurity problems, such as ALPS \cite{ALPS}, W2dynamics\cite{W2dynamics}, TRIQS \cite{triqs} CT-HYB \cite{ct-hyb} and CT-SEG \cite{ct-seg} (which can be used together with TRIQS DFT Tools \cite{DFTTools} or Solid DMFT \cite{solid_dmft} framework), comDMFT \cite{comDMFT}, etc., the packages, described in this paper, namely 
\begin{enumerate}[label=\roman*]
    \item \texttt{Hubb\_DMFT} (Single-Band Hubbard Model DMFT) — an optimized code for the standard single-orbital physics within the Hubbard model, 
    \item \texttt{Wan2mb\_DMFT} (Wannier to Many-Body DMFT) — a flexible automated tool designed for realistic materials calculations using multi-band Wannier Hamiltonians, designed for DMFT modeling of strongly correlated systems,

\end{enumerate}
    provide easy to use possibility of solving the DMFT impurity problem (together with the possibility of calculating the respective vertex functions), being relatively fast and accurate.
These programs wrap the modified iQIST core v. 0.7.0 \cite{iqist,iQist1} into a fully automated, self-consistent DMFT loop that natively interfaces with tight-binding models and inherits improved estimators for the self-energy and vertices, which are present in the iQIST package and allow for a reduction of strongly stochastic noise. The multi-orbital version features built-in parsers for standard Wannier Hamiltonian outputs (such as \texttt{wannier90\_hr.dat}), manages orbital-dependent Coulomb interactions (Kana\-mori / Slater), and handles the automated double-counting corrections required for realistic material simulations.

    The \texttt{Hubb\_DMFT} program was first developed in spite of the study of the extended Hubbard model in Ref. \cite{EDMFT}, the \texttt{Wan2mb\_DMFT} was developed first in spite of the study of exchange interactions in Ref. \cite{Our}.

\section{Software Features}
Both \texttt{Hubb\_DMFT} and \texttt{Wan2mb\_DMFT} are based on the internally integrated codebase of \mbox{iQIST} (intelligent Quantum Impurity Solver Toolkit) \cite{iqist}, developed in modern {Fortran}, leveraging its native efficiency for heavy numerical linear algebra. In accordance with open-source licensing, we have introduced modifications in iQIST solver, which allow treatment of the realistic single- and multiorbital problems. In the single-orbital case we implement in addition to previous realization \cite{iqist} the following features:
\begin{enumerate}[label=\roman*]
    \item The adjustment of the chemical potential to fix the electron density.
    \item{Calculation of the 3-point boson-fermion vertices (together with the 4-point vertices), which are necessary for some non-local extensions\cite{EDMFT,Review,DGA,abinitioDGA,DF,DB,TRILEX,DMF2RG,DTRILEX}}.
    \item{Separate cycle of CT-QMC calculation of the vertices after the self-consistent cycle. Possibility of specifying different numbers of bosonic and fermionic frequencies for vertex calculations}.
    \item{The accumulation of QMC data for the vertex from multiple cores is improved to make calculations with many bosonic and fermionic frequencies possible}.    
    \item{Possibility to consider arbitrary density of states; in particular, the density of states for the two-dimensional $t$-$t'$ Hubbard model is provided.}
    \item{E-DMFT self-consistent cycle for the nonlocal interaction $V_{\mathbf q}$.}
\end{enumerate}
For the multi-band Hubbard model in addition to i-iv above, we implement:
\begin{enumerate}[label=\roman*]
\setcounter{enumi}{7}
\item{Import of Wannier90 band structure.}
\item{Diagonalization of the crystal field of correlated atoms.}
\item{Double counting correction, implemented in around mean field (AMF), fully localized limit (FLL), and nominal filling.}
\end{enumerate}

\section{Theoretical Background and Method}
For the single-band model we consider the Hubbard Hamiltonian
\begin{equation}
    H = \sum_{ij, \sigma} t_{ij} c_{i\sigma}^\dagger c_{j\sigma} + U \sum_{i} n_{i\uparrow} n_{i\downarrow}+(1/2) \sum_{\mathbf q} V_{\mathbf q} n_{\mathbf q}n_{-{\mathbf q}}.
\end{equation}
where $t_{ij}$ represents the hopping matrix elements (determining the density of states $\rho(\epsilon)$), $U$ and $V_{\mathbf q}$ are the parameters of Coulomb interaction.

For the multi-band model we consider the Hubbard model is considered with the density-density interaction:
\begin{equation}
    H_{\text{DMFT}} = H_{\text{DFT}}^{\text{WF}} + \sum_i H_{\rm int}[c_{im},c_{im}^+] - \sum_{im} M_{{\text{DC}},i} n_{im},
\end{equation}
where $m,m'$ are the spin and orbital indices,
\begin{align}
    H_{\text{DFT}}^{\text{WF}} = \sum_{\mathbf{k}, \lambda\lambda', \sigma} H_{\mathbf{k}}^{\lambda\lambda'} c_{\mathbf{k}\lambda\sigma}^{+} c_{\mathbf{k}\lambda'\sigma},\\
    H_{\rm int}=\frac{1}{2} \sum_{mm'} U_{mm'} n_{im} n_{im'}\label{Hint}
\end{align}
are the Hamiltonians of non-interacting electrons and the interaction. The matrix $U_{mm'}$ is parameterized by the Slater parameters $U = F^0$, $J_H = (F^2 + F^4)/14$, $M_{{\rm DC},i}$ is the double counting correction. 
The program implements the  around mean field (AMF) form of the correction:
\begin{equation}
    M_{{\text{DC}},i} = \langle n_{id} \rangle [U(2n_{\text{orb}} - 1) - J_H(n_{\text{orb}} - 1)] / (2n_{\text{orb}}),
\end{equation}
the fully localized limit (FLL) form:
\begin{equation}
    M_{{\text{DC}},i} = U(\langle n_{id} \rangle - 1/2) - J_H(\langle n_{id} \rangle - 1)/2,
    \label{FLL}
\end{equation}
and nominal form of the double counting with $\langle n_{id} \rangle$ in Eq. (\ref{FLL}) replaced by the nominal filling $n_\mathrm{nom}$.

The fundamentals of the DMFT method are outlined in the review \cite{DMFT}. The method maps the original Hubbard model onto a set of impurity problems:
\begin{align}
    S_{\text{imp}} &= -\int_0^\beta d\tau \int_0^\beta d\tau' \sum_{m\sigma} c_{m}^+(\tau) [\mathcal{G}_m^0(\tau-\tau')]^{-1} c_{m}(\tau') \notag \\&+  \int_0^\beta d\tau  H_{\rm int}[c_{m},c^+_{m}],
\end{align}
where $H_{\rm int}[c,c^+]=U  n_{\uparrow} n_{\downarrow}+(1/2) \sum_{\omega} v_{\omega} n_{\omega}n_{-{\omega}}$ for the single-band model ($v_{\omega}$ is the bosonic bath) and $H_{\rm int}[c,c^+]$ is given by the Eq. (\ref{Hint}) for multi-orbital Hamiltonian.

The self-consistent DMFT loop implemented in both programs proceeds via the following steps:
\begin{enumerate}
    \item \textbf{Impurity Solver:} Calculate the local Green's function $G_{\text{local}}(\tau)$ from the effective bath hybridization $\Delta(\tau)$ using the internal \text{CT-HYB} algorithm.
    \item \textbf{Dyson Equation:} Extract the local electronic self-energy: $\Sigma(i\omega_n) = G_0^{-1}(i\omega_n) - G_{\text{local}}^{-1}(i\omega_n)$.
    \item \textbf{Lattice Hilbert Transform:} Calculate the updated local Green's function via Brillouin zone integration over the $\mathbf{k}$-mesh:
    \begin{equation}
        G_{\text{local}}(i\omega_n) = \sum_{\mathbf{k}} \left[ (i\omega_n + \mu)\mathbb{I} - H_{0}(\mathbf{k}) - \Sigma(i\omega_n) \right]^{-1}.
    \end{equation}        
    \item \textbf{Update Bath:} Compute the new hybridization function $\Delta^{\text{new}}(i\omega_n)$ and repeat until convergence is reached.
        For the E-DMFT cycle of the single-band Hubbard model, the bosonic bath $v(i\omega_n)$ is also adjusted according to 
            \begin{equation}
        \chi_{\text{local}}(i\omega_n) =\frac{\Pi(i\omega_n)}{1-v(i\omega_n)\Pi(i\omega_n)}= \sum_{\mathbf{q}} \frac{\Pi(i\omega_n)}{1-V_{\mathbf q}\Pi(i\omega_n)}.
    \end{equation}
    \item \textbf{Mix old and new solution}: For the single-band model (\texttt{Hubb\_DMFT}) Broyden method of mixing is used, while linear mixing is implemented in \texttt{Wan2mb\_DMFT}.

\end{enumerate}

\section{Input and output format}

\subsection{Input Files Preparation}

\subsubsection{The file \texttt{\rm solver.ctqmc.in}.}

The file \texttt{solver.ctqmc.in} should be prepared according to standard iQIST input format \cite{iqist,iQist1} with the following additional parameters:
\begin{tabbing}
   Mom $p$ \qquad \= \kill 
    
    {part:} \> if nonzero used to specify the fixed density \\[3pt]
    {isvrt:}     \> \parbox[t]{11.5cm} {6-th bit set to 1 (i.e. adding $2^5=32$) corresponds to  the three-point vertex calculation. This calculation also requires evaluation of frequency dependent bosonic susceptibility to be switched on (5-th bit of \texttt{issus} should be set to 1, i.e. $2^4=16$ should be added to \texttt{issus}) and works presently only with the standard vertex evaluation, \texttt{isort}=1. Second or third bit of \texttt{isvrt} set to 1 (i.e. adding $2^1=2$ or $2^2=4$) corresponds (as in it is originally designed in iQIST) to the evaluation of the 4-point vertex, parameterized in the particle-hole channel.}\\[6pt]
    {nffrq1:} \> \parbox[t]{11.5cm}{specifies the number of fermionic frequencies in the vertex calculation (put to zero if the vertices are not needed)} \\[3pt]
    {nbfrq1:}     \> \parbox[t]{11.5cm}{specifies the number of bosonic frequencies in the vertex calculation.}
\end{tabbing}

\subsubsection{Other options in \texttt{\rm solver.ctqmc.in} for the single-orbital model}

\begin{tabbing}
   Mom $p$ \qquad \= \kill 
    
    {t1:} \> \parbox[t]{11.5cm}
    {specifies $t'/t$ for the square lattice density of states (DOS) with nearest ($t$) and next-nearest neighbors ($t'$) hopping parameters. The parameter \texttt{t1} can be also used for the other types of DOS. The shape of the DOS is programmed in the files src/ctqmc/nrcissus/ctqmc\_dmft.f90, ctqmc\_stream.f90 (the latter contains DOS-specific initialization of the hybridization function).} 
\end{tabbing}

\subsubsection{\texttt{\rm solver.hybXX.in, solver.sgmXX.in, solver.mune.in} files} 
These files are optional and should be prepared in case the calculation starts from the results of one of the previous iterations. These files can be copied from the respective \texttt{.dat} files of that iteration, XX is not present for the single-band \texttt{Hubb\_DMFT} calculation and denotes the number of DMFT impurity (` 1' for the first impurity, and so on) for \texttt{Wan2mb\_DMFT} version. 

\subsubsection{Other files and options for multi-orbital models}
\vspace{0.3cm}
\begin{enumerate}[topsep=3pt]

    \item \texttt{solver.ctqmc.in} \\[5pt]
    Options related to multi-orbital model simulation are described below:
\begin{tabbing}
   Mom $p$ \qquad \= \kill
   
    {nat:} \> \parbox[t]{10.5cm}{number of the considered orbital sets (see the description of the content of the file imp.in below),}  \\
        {idc:} \> \parbox[t]{10.5cm}{double-counting (DC) correction type: 0 - AMF, 1 - FLL, 2 - nominal,}\\
        {nnom:} \> the filling, used in nominal double counting, \\
        {part:} \> \parbox[t]{10.5cm}{number of valence electrons in the Wannier orbitals per unit cell (must be present).}

\end{tabbing}

In addition, the following files should be prepared:

    \item \texttt{Recip.in} - Used reciprocal lattice vectors.
Format:

b1x b1y b1z

b2x b2y b2z

b3x b3y b3z

The values can be copied from the Wannier90 \texttt{*.wout} file.

    \item \texttt{imp.in}  - List of considered orbital sets. Format:

nat lines

\texttt{ic} \texttt{kind} \texttt{U} \texttt{J$_H$} \texttt{n$_{\rm orb}$} \vspace{.1cm}\\ 
    where:
    \begin{itemize}[label=,itemsep=0pt,topsep=-1pt]
        \item \texttt{ic} - index of the first orbital of the orbital set in the Wannier  Hamiltonian orbital space (counted from 0).
        \item \texttt{kind} - row number of the orbital set to which this orbital set is equivalent in establishing self-energy equivalence (after rotation of the crystal field). To consider the current orbital set as an impurity in DMFT, put this value equal to the current row number. If the orbital set is not aimed to be considered within QMC (but its occupation numbers are displayed), put \texttt{kind}=0. 
        \item \texttt{U, J$_H$} - Coulomb interaction magnitude at this orbital set in Slater parameterization ($\texttt{U}=F^0$, $\texttt{J}=(F^2+F^4)/14$, ignored if the orbital set is specified being equivalent to another orbital set).
        \item \texttt{n$_{\rm orb}$} - number of orbitals in the current orbital set.
    \end{itemize}

    After specifying all orbitals, which are considered within QMC (which equivalence is set by their \texttt{kind}s), the user can also list the other (\texttt{kind}=0) uncorrelated orbitals ($U=J=0$), or the orbitals considered within Hartree-Fock approximation. 

    \item \texttt{dos.dat}. Format of the first line (which is only used):

[a string of 42 arbitrary characters][xxxxx.xxx]

    \texttt{xxxxx.xxx} - the DFT chemical potential. If Quantum Espresso is used, the corresponding file can be copied from the Quantum Espresso output data.

    \item \texttt{hr.dat} \\
    The real-space Hamiltonian generated by Wannier90.

    \item \texttt{kpoints.dat}. Parameters of the momentum grid and $k$-points to write Hamiltonian eigenvalues along high-symmetry directions. Format:\vspace{.25cm}\\
\texttt{nx} \texttt{ny} \texttt{nz}
(number of grid splits in k-space for each direction, does not necessarily have to match the grid used in Wannier90) followed by lines

\texttt{Label} \texttt{ikx} \texttt{iky} \texttt{ikz}

    where \texttt{Label} is any character identifying the k-point (can be a space), and \texttt{ikx,iky,ikz} are integer (not necessarily positive) coordinate indices of the $(k_x,k_y,k_z)$ point in terms of the reciprocal lattice vectors: \texttt{ikx=[$k_x$*nkx]}, etc.

\end{enumerate}

        \subsection{Output files}

The developed programs use iQIST structure of the output files \cite{iQist1} with the following additions/changes:

\begin{enumerate}[topsep=3pt]

\item{Each file name contains after the standard name the number of DMFT iteration. For the iteration, corresponding to vertex calculations, the number 999 is used.}

\item For Wan2mb version each output file has the number of DMFT impurity in the end of the file name. 

\item The file \texttt{solver.mune.dat} contains the chemical potential values (and double counting potentials for \texttt{Wan2mb\_DMFT}) at each iteration.

\item The file \texttt{solver.v4phXX.dat} contains fermion interaction vertices 
\begin{align}
\Gamma_{\nu\nu'\omega}^{mm'}&=G_{\nu,m}^{-1}G_{\nu+\omega,m}^{-1}G_{\nu',m'}^{-1}G_{\nu'-\omega,m'}^{-1}\left[\langle c^+_{\nu,m} c_{\nu+\omega,m} c^+_{\nu',m'} c_{\nu'-\omega,m'}\rangle\right.\\
&-\left.\langle c^+_{\nu,m} c_{\nu,m}\rangle \langle c^+_{\nu',m'} c_{\nu'm'}\rangle\delta_{\omega,0}+\langle c^+_{\nu,m} c_{\nu,m}\rangle \langle c^+_{\nu+\omega,m} c_{\nu+\omega,m}\rangle\delta_{\nu+\omega,\nu'}\delta_{mm'}\right]\notag
\end{align}
for the impurity site XX (the number is not specified in the single-orbital case), where $\nu,\nu'$ are the fermionic frequencies, $\omega$ is the bosonic frequency, and $m,m'$ are the spin-orbital indices. The format of the file is 

\# flvr1:     $m$

\# flvr2:     $m'$

\# nbfrq:     $n_\omega$

...

$n_\nu$ $n_{\nu'}$ Re$\Gamma$ Im$\Gamma$ Re$\Gamma^{\rm imrv}$ Im$\Gamma^{\rm imrv}$

...

where $\Gamma^{\rm imrv}$ is estimated using the improved estimators.

\item The file \texttt{solver.v43phXX.dat} contains triangular vertices 
\begin{equation}
\Gamma_{\nu\omega}^{mm'}=G_{\nu,m}^{-1}G_{\nu+\omega,m}^{-1}\left[\langle c^+_{\nu,m} c_{\nu+\omega,m} n_{\omega,m'}\rangle-\langle c^+_{\nu,m} c_{\nu,m}\rangle \langle n_{\omega,m'}\rangle\delta_{\omega,0}\right], 
\end{equation}
for the impurity XX site XX, where $\nu$ is the fermionic frequency, $\omega$ is bosonic frequency, $m,m'$ are spin-orbital indexes, $n_{\omega,m}$ is the Fourier transform of time-dependent electron density at the spin-orbital $m$. The format of the file is similar to that for the full vertex 

\# flvr1:     $m$

\# flvr2:     $m'$

\# nbfrq:     $n_\omega$

...

$n_\nu$ Re$\Gamma$ Im$\Gamma$ Re$\Gamma^{\rm imrv}$ Im$\Gamma^{\rm imrv}$

...

\end{enumerate}

In addition, the multi-orbital version Wan2mb writes the following files:

\begin{enumerate}[topsep=3pt]
\setcounter{enumi}{5}
\item \texttt{solver.rotm.dat} contains rotation matrix of the Hamiltonian, which transforms the crystal field to the diagonal form.
\item \texttt{bands\_ham.dat} contains eigenvalues of the Hamiltonian along the path, chosen in kpoints.dat.
\item \texttt{solver.umat.dat} contains interaction matrix $U_{mm'}$ for equal (diagonal blocks) and non-equal (off diagonal blocks) spins.
\item \texttt{solver.g0.dat} contains initial local Green's function for each set of orbitals.
\item \texttt{solver.eimp.dat} contains the atomic levels of each DMFT impurity.
\end{enumerate}


\subsection{Hybrid Parallelization Strategy}
To meet the massive computational demands of multi-orbital DMFT, the codes inherit a multi-layered hybrid parallelization scheme of iQIST using \textbf{MPI} and \textbf{OpenMP}:
\begin{itemize}
    \item \textbf{Inter-node (MPI):} Independent Monte Carlo Markov chains are distributed across multiple compute nodes via MPI. In \texttt{Wan2mb\_DMFT}, MPI is also utilized to parallelize the costly Brillouin zone integration over the $\mathbf{k}$-mesh during the lattice Dyson equation step.
    \item \textbf{Intra-node (OpenMP):} Shared-memory parallelization via OpenMP can be employed within each node to accelerate cycles over frequencies and orbital indices in the vertex calculations.
\end{itemize}

\section{Compiling, Test Cases and Verification}

To compile the program, it is sufficient to execute \texttt{cd build ; make all} or
run the script \texttt{compile.sh} in the respective example directories. Adjust the file \texttt{build/make.sys} appropriately. To run, execute \texttt{src/ctqmc/\allowbreak narcissus/ctqmc}. The example of running SLURM script is present in \texttt{run.sh} file in example directories.

To validate the accuracy, performance, and versatility of the presented packages, we provide two distinct benchmark cases, bridging a foundational model of cuprates and realistic, state-of-the-art materials simulation.

\begin{figure}[t]
\centering
\includegraphics[width=0.48\textwidth]{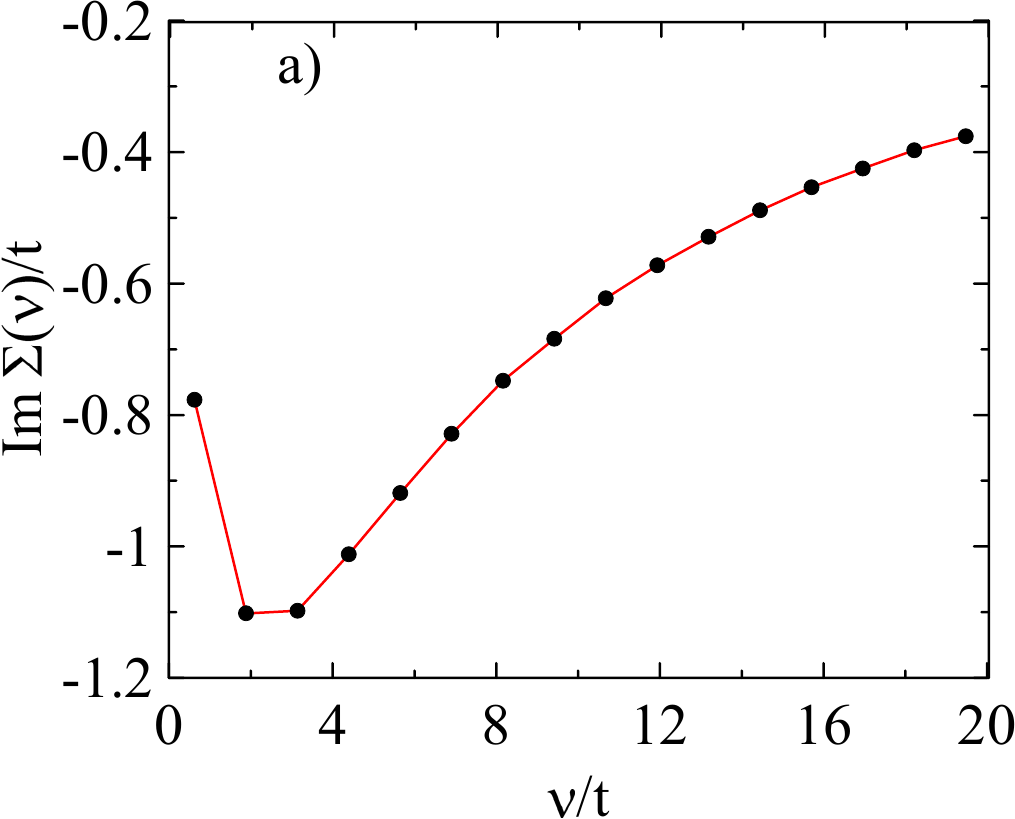}
\includegraphics[width=0.46\textwidth]{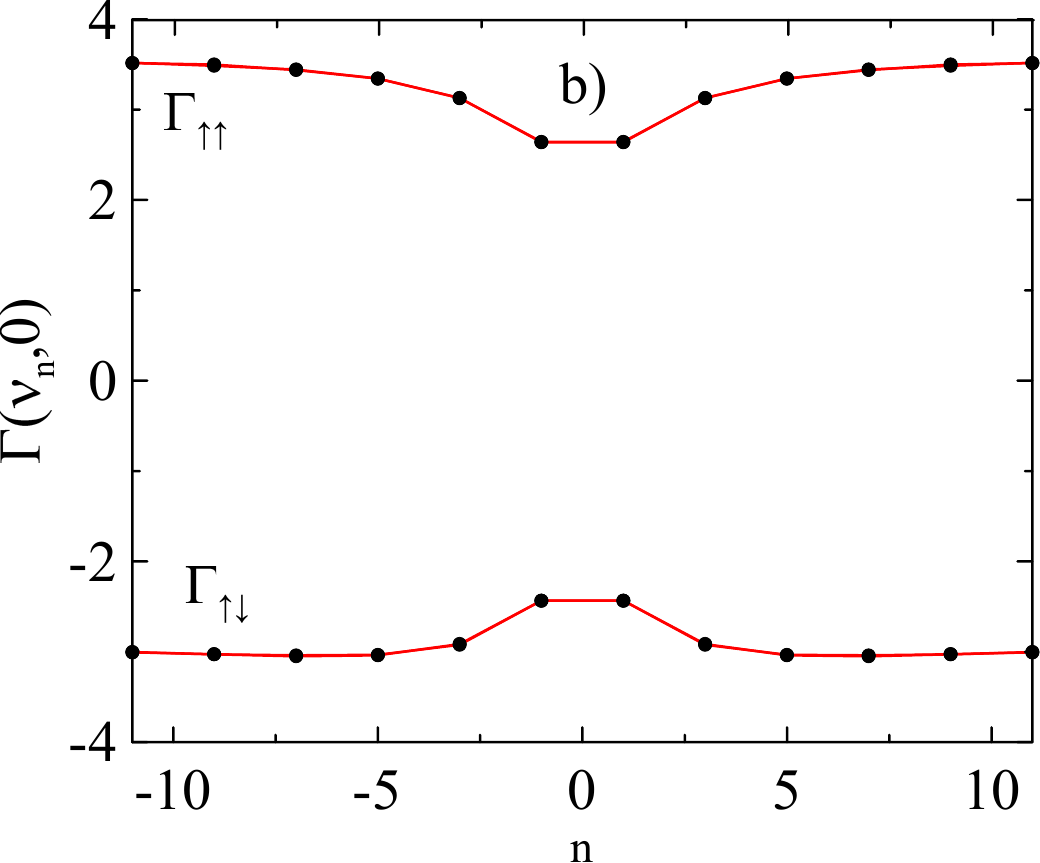}
\caption{Calculated imaginary part of the Matsubara self-energy $\Sigma(i\nu_n)$ (a) and triangular vertices $\Gamma^{\sigma\sigma'}_{\nu,0}$ (b) for the 2D Hubbard model at $t'/t = -0.3$, $U=5.6t$, $n=0.96$ at $T=0.2t$.}
\label{fig:one_band}
\end{figure}

\subsection{{\rm\texttt{Hubb\_DMFT}}: Two-Dimensional Hubbard Model with Next-Nearest Neighbour Hopping}
We consider an application of \texttt{Hubb\_DMFT} on a square lattice, including both nearest-neighbor ($t$) and next-nearest-neighbor ($t'$) hopping matrix elements. This model is a standard paradigm for high-temperature cuprate superconductors. We set $U=5.6t$, a ratio of $t'/t = -0.3$ at a representative temperature $T/t = 0.2$ and filling $n=0.96$. The calculated self-energy (Fig. \ref{fig:one_band}a) $\Sigma(i\omega_n)$ agrees with that obtained previously in Ref. \cite{Kugler2025}. The respective triangular vertices are shown in Fig. \ref{fig:one_band}b.



\subsection{{\rm \texttt{Wan2mb\_DMFT}}: Two-Dimensional Ferromagnetic Van der Waals Material $\text{CrTe}_2$}

\begin{figure}[h]\centering
\includegraphics[width=0.47\textwidth]{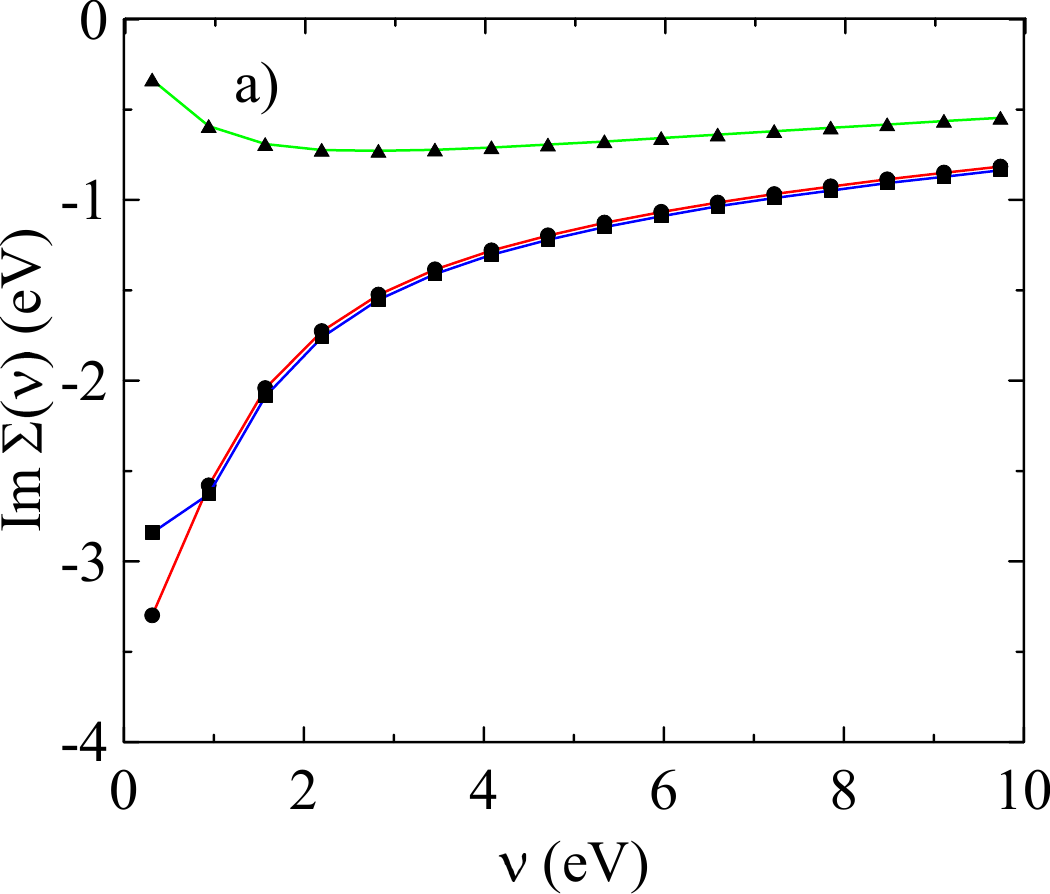}
\includegraphics[width=0.49\textwidth]{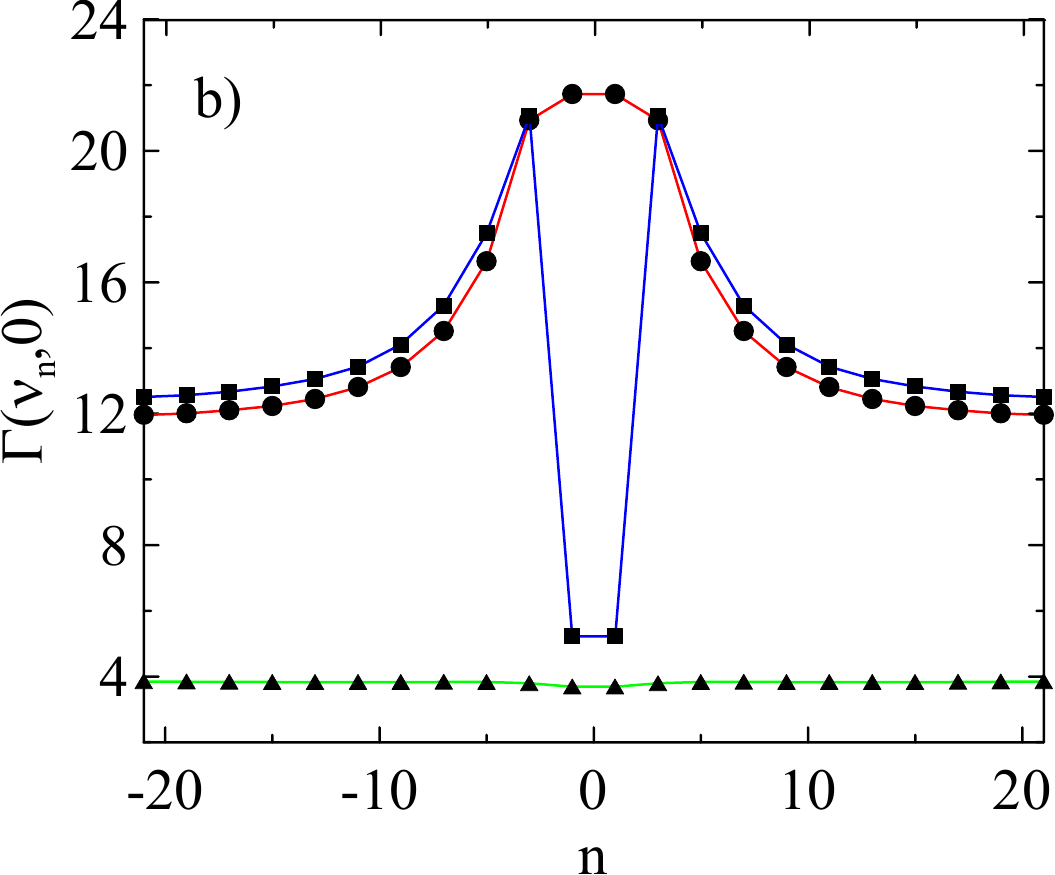}
\caption{Calculated imaginary part of the Matsubara self-energy $\Sigma(i\nu_n)$ (a) and triangular vertices $\Gamma^{m\uparrow,m\uparrow}_{\nu,0}$ (b) for a monolayer of $\text{CrTe}_2$ at $\beta=10$~eV$^{-1}$, capturing orbital dependence of the calculated quantities.}\label{fig:crte2}\end{figure}

To highlight the capability of \texttt{Wan2mb\_DMFT} in handling modern, low-dimensional correlated systems, we simulate a monolayer of the transition metal dichalcogenide $\text{CrTe}_2$. The Wannier basis consists of the correlated $\text{Cr}$-$3d$ orbitals (states 1..5) coupled with the $\text{Te}$-$5p$ ligand states (states 6..11) to explicitly capture the $p$-$d$ hybridization. The Wannier90 Hamiltonian is generated and present in the file \texttt{hr.dat}. A density-density interaction is applied to the $\text{Cr}$ site with $U=2.8$~eV and $J=0.9$~eV. The respective content of the \texttt{imp.in} file is shown in Fig. \ref{Fig:Impin}. The results of the calculations are shown in Fig. \ref{fig:crte2}. The solver captures the strong orbital-dependent correlation effects within the $a_{1g}$, $a_{2g}$ and $e_g$ manifolds of chromium. The results provide a clear picture of orbital-dependent electronic correlations.
\begin{figure}[t]
\centering
\label{tab:data_table}
\setlength{\tabcolsep}{12pt} 
\begin{tabular}{|ccccc|}
\hline
\rule{0pt}{4ex} 
0                 & 1                 & 2.8               & 0.9               & 5                 \\
5                 & 0                 & 0.0               & 0.0               & 3                 \\
8                 & 0                 & 0.0               & 0.0               & 3                 \\[1.5ex] 
\hline
\end{tabular}
\caption{Content of the imp.in file for CrTe$_2$}
\label{Fig:Impin}
\end{figure}

\section*{Acknowledgments}The authors acknowledge the developers of the \texttt{iQIST} package, whose foundational open-source code served as the initial codebase for the core impurity solver engine. The author is also grateful to I. A. Goremykin, T. B. Mazitov, E. Agapov, and I. S. Dedov who helped to test the software at various stages of its development.

\end{document}